\documentclass[preprint]{aastex}
\shorttitle{Long-period variability in {\it o} Ceti}
\shortauthors{Templeton \& Karovska}
\begin{document}
\title{Long-Period Variability in {\it o} Ceti}
\author{Matthew R. Templeton}
\affil{American Association of Variable Star Observers, 49 Bay State Road, Cambridge, MA 02138}
\author{Margarita Karovska}
\affil{Harvard-Smithsonian Center for Astrophysics, 60 Garden Street, Cambridge, MA 02138}

\begin{abstract}
We carried out a new and sensitive search for long-period variability in the
prototype of the Mira class of long-period pulsating variables, {\it o}
Ceti (Mira A), the closest and brightest Mira variable.   We conducted this
search using an unbroken light curve from 1902 to the present assembled from
the visual data archives of five major variable star observing organizations
from around the world.  We applied several time-series analysis
techniques to search for two specific kinds of variability: long secondary
periods (LSPs) longer
than the dominant pulsation period of $\sim 333$ days, and long-term
period variation in the dominant pulsation period itself.  
The data quality is sufficient to detect coherent periodic variations 
with photometric amplitudes of 0.05 mag or less.  We do not
find evidence for coherent LSPs in {\it o} Ceti to a limit of 0.1
mag, where the amplitude limit is set by intrinsic, stochastic,
low-frequency variability of approximately 0.1 mag.  We marginally
detect a slight modulation of the pulsation period similar in timescale as
that observed in the Miras with meandering periods, but with a much lower
period amplitude of $\pm 2$ days.  However, we do find clear evidence of a
low-frequency power law component in the Fourier spectrum of {\it o} Ceti's
long-term light curve.  The amplitude of this stochastic
variability is approximately 0.1 mag at a period of 1000 days,
and it exhibits a turnover for periods longer than this.  This spectrum is 
similar to the red noise spectra observed in red supergiants.
\end{abstract}


\section{Introduction}
The bright variable {\it o} Ceti (Mira A) is the class prototype of the 
{\it Mira variables}, which are pulsating asymptotic giant branch (AGB) stars
with periods of hundreds of days, and visual light amplitudes of between 2.5 
and 10 mag.  {\it o} Ceti, discovered
in 1596, is also the brightest member of its class,
with visual maxima routinely reaching 3rd magnitude.  {\it o}
Ceti is also one of the closest Mira variables to us, and because of its
large size (over $500 R_{\odot}$) was resolved with the 
Hubble Space Telescope and interferometric techniques
(e.g. \citet{Karovska91,Wilson92,Haniff95,Karovska97,Woodruff08}).
Because of its brightness, its
proximity, and its long observational history, {\it o} Ceti provides
an excellent laboratory for studying the Mira variables generally.
{\it o} Ceti is also a symbiotic binary, as it has a companion --
VZ Ceti (Mira B) -- at a distance of approximately 70 AU
(e.g \citet{Karovska92,Karovska05,Ireland2007})
which is accreting mass from Mira A's substantial wind.

Pulsating AGB stars generally show very irregular pulsation behavior.  The 
Mira variables are an exception to the rule among AGB stars, but even so they
are generally not stable in period at a level of less than 1\% from
cycle-to-cycle, and their maxima, minima, and light curve shape often vary
significantly from cycle-to-cycle as well.  A small 
fraction have large excursions in period, amplitude, and mean light over the 
course of their observational histories, the causes of which are not known. 
A small fraction ($\sim 10\%$) exhibit clearly
measurable period variations over time scales of decades; these 
stars were dubbed {\it meandering Miras} by \citet{ZB02} and tend to have 
nonmonotonic period variations of order $5\%-10\%$ of their pulsation period.
A smaller fraction still ($\sim 1\%-2\%$) have exhibited continuously increasing
or decreasing periods with changes exceeding $\sim 10\%$, and may be in the 
midst of {\it thermal pulses} \citep{WZ81}.  Large period changes, though rare, have
been well documented in a few stars since the earliest work of 
\citet{SC37}, and centuries-long historical light curves for a few
Miras show significant changes over time (\citet{Sterken99,PA99,ZBM02}), even
if periods are relatively stable at present.  \citet{TMW05} used the 
American Association of Variable Star Observers (AAVSO) data archives to 
search for large, long-term period changes in 547 Mira variables, including
{\it o} Ceti itself.  Although {\it o} Ceti was not flagged as having a 
large period change, the significance criteria used in that paper were very 
high, and did not rule out period changes at the level of a few percent. More
subtle period changes including the seemingly random cycle-to-cycle
ones may be of astrophysical interest.

{\it o} Ceti itself may exhibit long-term variations in pulsation behavior
as well.  \citet{BM97} used the visual data from the AAVSO data archives,
and found long periods (from 600 to over 1500 days) with marginal statistical
significance.  The detections were not of high statistical significance
(primarily due to broadband, low-frequency power), but this raises an 
interesting question about long periods in {\it o} Ceti and in the Miras 
generally.  There have been reports of other long-term variations in the
Mira AB system in the past \citep{Joy54,YM77,Baize80,Karovska92,WK06}.

Long periods and long-period variations are known to occur in AGB pulsators,
though the physical processes that generate them are not understood, and there
is no conclusive evidence of long multiperiods in Miras.  \citet{Hinkle02}
used spectroscopy to study a sample of nine AGB variables with long secondary
periods (LSPs), and found six of the nine stars showed radial velocity 
variations that confirmed the photmetric secondary periods but are interpreted
as {\it not} being caused by binarity.  The one Mira variable in
their sample with a suspected LSP (SV And) showed no
secondary radial velocity variations.
\citet{WOK04} also used spectroscopy to study three AGB stars known to have
secondary periods in an attempt to determine the origin of this behavior.
They found that the data support none of the suspected origins of these
periods (radial or nonradial pulsations, binarity, or stellar activity and
spots) unambiguously; those authors suggest that low-$\ell, g+$ modes in the
radiative portion of the red giant are best supported by the data.  Among
the Milky Way population, \citet{PB03} showed that many pulsating
red giants (as many as 30\% of them) show LSPs on the order
of hundreds or thousands of days, in addition to their shorter periods of
tens to one hundred days.  
Recently, the spotlight has been placed on long-term secondary periodicity
in semiregular variables and pulsating giants observed in the
Large Magellanic Cloud (LMC). \citet{Wood99}
clearly showed that there also exists a well-populated sequence of red giant 
stars in the LMC period-luminosity diagram with secondary periods of several 
hundreds to 1000 days.  In both cases, the causes for these long periods are 
not clear, but
the suggestions of binarity, and nonradial, strange-mode, or dust-driven 
pulsation were invoked.  Long-period variations clearly exist in red giant
AGB pulsators, but the physical picture of why they occur is as yet unclear.

We have undertaken a new study of the long-term light curve of {\it o} Ceti 
with the aim of exploring the long-term variability of the class
prototype of the Mira variables.  It is critical, given the huge
amount of observational time devoted to this object, that the long-term
observational record be sensitively and self-consistently analyzed to assess
whether long-term variations truly do occur in {\it o} Ceti, and how they
manifest themselves.  In Section 2, we present a newly
compiled light curve for {\it o} Ceti drawn from all available archives of
visual observations, and discuss techniques and strategies that we used to 
search for long-term variations in the data.  In Section 3, we present our
results, discussing the nature of the observed variabilities and placing 
statistical limits on the likelihood of coherent secondary periods and
long-term variability.  In Section 4, we present a discussion of the results.

\section{Data \& Analysis}

The primary data set of {\it o} Ceti used in this study is composed entirely
of visual magnitude estimates made by several hundred individual observers 
since 1902.  The majority of these data were taken from the AAVSO 
{\it International Database}.  The AAVSO International Database already 
includes 
observations from the {\it Association Francaise des Observateurs d'Etoiles 
Variables} (AFOEV).  To these data, we added observations not already 
submitted to the AAVSO archives from three other organizations: the 
{\it British Astronomical Association-Variable Star Section} (BAAVSS), 
the {\it Royal 
Astronomical Society of New Zealand} (RASNZ), and the {\it Variable Star 
Observers League of Japan} (VSOLJ).  The AAVSO data have been checked for
transcription and other errors with a {\it validation} procedure described in
\citet{valid}.  Data from the three other organizations were not validated
during the AAVSO data validation project, but were checked for transcription 
or keypunch errors for the present project, and all duplicate 
and highly discrepant points (more than 1 mag away from
the mean) were removed prior to analysis.  The resulting visual light curve 
covers the time period of JD 2,415,998.5 (1902 September 06) to JD 2,453,883.9 
(2006 May 28), and is the first light curve assembled by combining data from
all major variable star data archives worldwide.

We formed 10 day averages of the visual observations to reduce the scatter 
from both random errors and systematic differences between observers, and 
to ensure that the analyses are not temporally biased toward the better 
sampled parts of the light curve.  Averaging increases the signal-to-noise
of the visual data at the minor cost of reduced temporal sensitivity; since 
all periods of interest are larger than 20 days, the 10 day averaging has 
minimal impact on our analysis.  To assess the internal photometric 
consistency of the data, we measured the standard deviation of the visual
magnitude estimates in each 10 day bin; this quantity averaged 0.2 mag
at both maximum and minimum, and about 0.4 mag during the rise and
decline phases.  The lower scatter during maximum and minimum is due to the
historically intensive coverage of {\it o} Ceti during maximum and minimum 
phases, and to the fact that the star changes more over a 10 day span during
the rise and decay phases than during maximum and minimum.  The magnitude
scatter during maximum and minimum is essentially identical, despite the
factor of a few hundred difference in the flux between the two phases.  This is
due both to the quality of the comparison star sequence for {\it o} Ceti, and
to the fact that the star is typically observed with larger-aperture 
telescopes at minimum compared to maximum.

\begin{figure}
\figurenum{1}
\label{figlight curve}
\epsscale{1.0}
\plotone{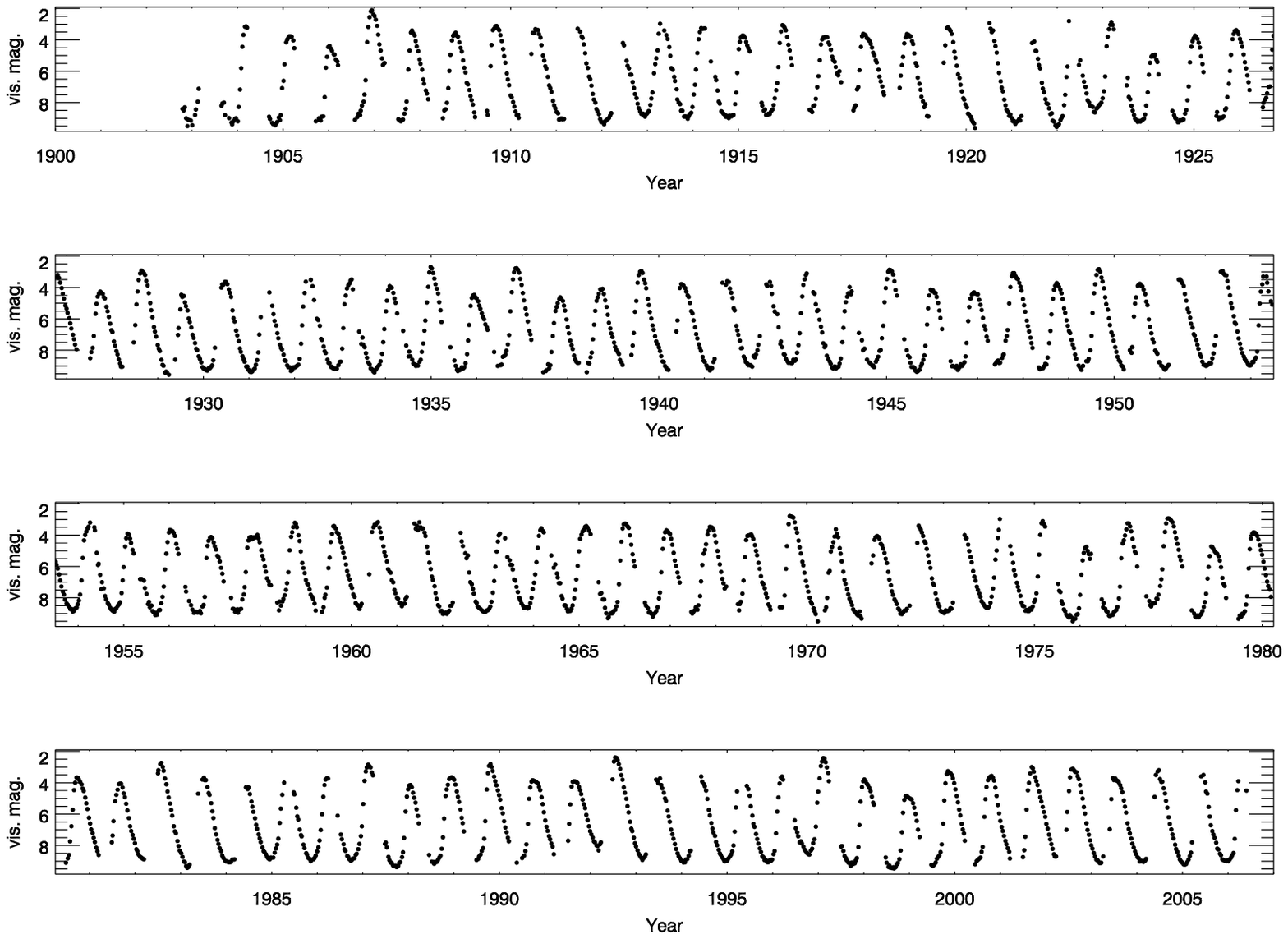}
\caption{Light curve of {\it o} Ceti, from 1902 to 2006.  Each point is
the average of 10 days' worth of visual observations.}
\end{figure}

The light curve shown in Figure \ref{figlight curve} contains all available data
from the five major variable star observing organizations, and is likely as
complete a record of the last century's history of {\it o} Ceti as has ever
been compiled.  Every cycle since 1902 has partial or complete coverage, with 
the only unavoidable exceptions being the annual gaps due to solar conjunction.
Many time-series analysis techniques are sensitive to gaps in data and so our
main motivation in compiling these data was the minimization of this source of
interference.  For our study, we focus our analysis only on these 105 years of
data.  There exist significant data prior to 1902, with good (but not complete)
coverage extending to the late 18th century, and very early observations
extending to Mira A's discovery in 1596.  This longer light curve was recently
collected from literature sources by E. Zsoldos \& G. Marschalk\'{o} and
analyzed for periodic amplitude variations \citep{M04}.  They have kindly
provided their light curve to us and we also analyze these data for 
LSPs and period variations as a check of our primary analysis.

The photometric behavior of {\it o} Ceti is complex, as it is in nearly all
Mira variables, and so we performed time-series analyses of the data using
several different techniques.  The use of multiple analysis techniques serves
both to more fully understand the behavior of the data and to provide a
consistency check, since all techniques should yield similar results to
within the statistical limitations of each.  We used three different classes
of tests: Fourier-based time-series analyses, phase-dispersion minimization
(PDM), and a Box-Jenkins autocorrelation model.  

Fourier-type algorithms used in this work include: the date-compensated 
discrete Fourier transform
(DCDFT) as derived by \citet{FM81} and implemented by \citet{Fosterdcdft};
the \citet{Roberts87} {\it clean} algorithm for which we developed our own
implementation; and the weighted wavelet Z transform, a time-frequency
analysis method developed and implemented by \citet{Fosterwwz}.  All of these
methods share a fundamental reliance on Fourier signal modeling, but have
different approaches to the problem.  The DCDFT is essentially a least-squares
model using periodic basis functions to find the best-fitting period,
amplitude, and phase; the times of the data points are explicitly taken into
account to allow uneven sampling.  The {\it clean} algorithm uses iterative 
peak finding and subtraction using the data window function to separate real
signals in the Fourier transform from spurious alias peaks caused by the 
annual data gaps.  Finally
the wavelet Z algorithm uses a standard Fourier fitting algorithm but
with the novelty of a sliding Gaussian weighting window to explore the change in
the data's signal content as a function of time.

Two other statistical tests were used.  First, we used the PDM algorithm of 
\citet{Stellingwerf78}
to test for long-term periodic behavior.  PDM folds the data on a set
of test periods, and bins the data for each test period to find that
with the lowest dispersion per bin, a sign of periodic behavior with that
period.  Second, we used an autoregressive integrated
moving average (ARIMA) model with a seasonal adjustment (i.e.
a SARIMA model) of the dominant pulsation period.  A \citet{BJ76} 
autocorrelation of the adjusted data was then done to test whether there 
were any other correlated timescales in the data.

\section{Results}

We sought to answer two questions about the long-term
behavior of {\it o} Ceti with our analyses: (1) how stable is the 
pulsation period over time, and (2) are there stable
LSPs present in the light curve?  The former
question is important because Miras are known to have strong cycle-to-cycle
variations in their light curves which statistically manifest themselves
as changes in period or amplitude, and it is important to determine whether 
observed period changes are simply due to random processes or something else
like evolutionary
or other long-term internal structural changes.  The latter question is clearly
also important because secondary periods are indicators of periodic processes
in or around the star, either internal pulsations or external modulations
such as interactions with a companion object.  The different time-series
techniques described in the preceding section can provide answers to 
different questions depending upon the type of analysis.  Fourier and
folding methods can most easily detect stable periods, but also provide
hints as to the stability of known periods.  Wavelet or time-frequency
methods like the wavelet Z transform can show the time evolution of
the spectrum, since it essentially performs a localized Fourier analysis
of short sections of the light curve.  Finally, autocorrelation can help
to uncover behavior which may not be strictly periodic, or which is
not present or in phase throughout the entire light curve.

During the course of our analysis, it became clear that one of the features of
the variability was a low-frequency power-law component, qualitatively similar
to that observed in other giant and supergiant pulsators by \citet{KSB06}.  A
third test that we performed was to quantify the nature of this
power law, and determine whether it quantitatively matched what was observed in
the red supergiants.  We discuss each of these three issues in the following
subsections.

\subsection{Pulsation period and period stability of {\it o} Ceti}

\begin{figure}
\figurenum{2}
\label{figcleanspec}
\epsscale{1.0}
\plotone{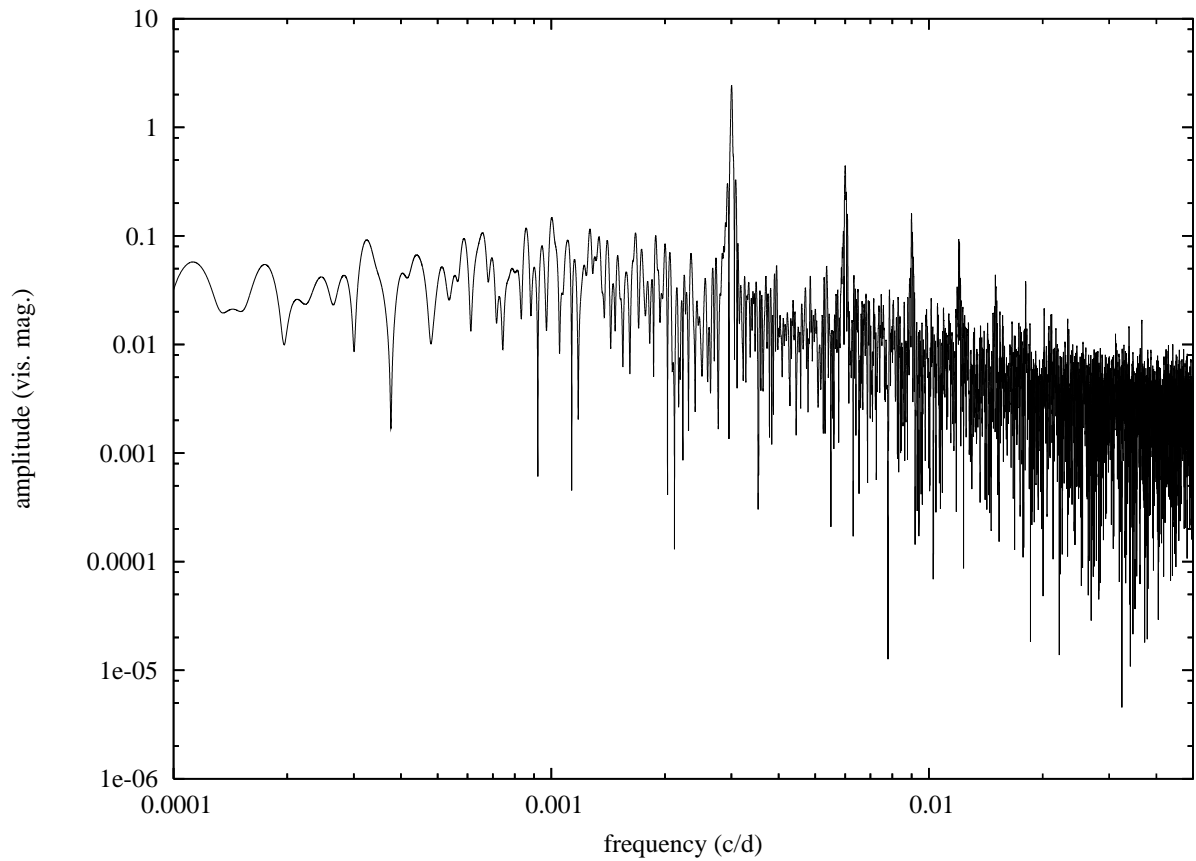}
\caption{Cleaned Fourier transform of the light curve of {\it o} Ceti
showing the dominant frequency of $0.003 c/d$ ($P$ = 333 days) and its
higher-frequency harmonics.  The spectrum is featureless other than
the low-frequency power-law continuum, and very weak peaks ($< 0.05$ mag)
at the lunar synodic and sidereal periods.}
\end{figure}

All time-series analysis methods yield the same dominant period of 
variability to within the calculational uncertainties of each; the
strongest period derived from the {\it clean} analysis is 
$333.09 \pm 0.04$ days (see Figure \ref{figcleanspec}).  All integer 
harmonics of this period through
$P_{0}/6$ (= 55.52 days) also appear in the Fourier spectrum, as do
the lunar sidereal and synodic periods, caused by a weak modulation
in sky brightness during the lunar cycle.  It is also apparent that the
dominant pulsation period is modulated, as the spectral peak is actually
a blend of multiple peaks, as are those of the Fourier harmonics; all of
the multiple peaks are contained within Lorentz- or Gaussian-shaped 
envelopes centered on the main period as was noted by \citet{KSB06}.
We used the Period04 \citep{P04} package to fit and prewhiten 
the data with the periods $P_{0}$ through $P_{0}/6$ noted above and 
found that subsequent peaks could not be reliably fit with any set of 
periods; allowing the periods of $P_{0}$ and its harmonics to 
vary along with those of additional peaks resulted in divergent
solutions unless unphysically large numbers of periods were used in the 
fit.  This indicates that the pulsation is changing with time; it is not
clear whether the period, amplitude, phase, or a combination of the three
are changing, but the time between individual light curve maxima and minima
are not constant to better than 4-5 days from cycle to cycle.

\begin{figure}
\figurenum{3}
\label{figwwz}
\epsscale{1.0}
\plotone{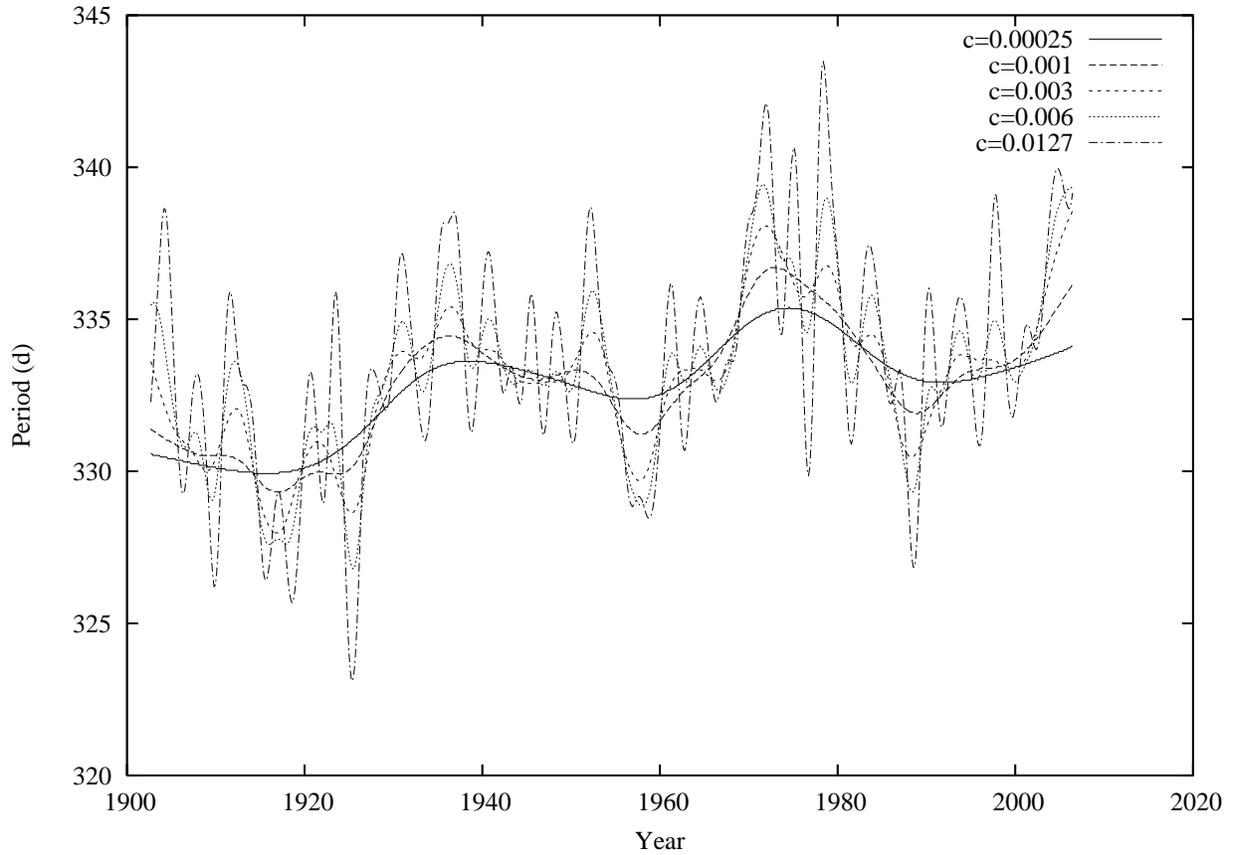}
\caption{Variation of the dominant period's centroid as a function of time,
measured with the weighted wavelet Z time-frequency algorithm \citep{Fosterwwz}.
Results are shown for several different widths of the weighting window,
ranging from $c=0.00025$ to $0.0127$.  For 
larger values of the window width parameter, $c$, the uncertainty in the 
period determination is much larger than the variation, but at the lowest 
values of $c$ the variation in period is on the order of the {\it maximum} 
uncertainty in the period.}
\end{figure}

We used the weighted wavelet Z transform \citep{Fosterwwz} to explore
how the pulsations varied with time, and whether the period was being
coherently modulated on a timescale measurable with the 105 year light
curve.  Figure \ref{figwwz} shows the peak period of {\it o} Ceti as
measured with several different window widths of the wavelet transform; 
the parameter $c$ is equal to $(2 \pi n_{c})^{-2}$, where $n_{c}$ is the
$e$-folding half-width of the weighting window in cycles.  As the width of 
the window is increased, the variations in period centroid become much less 
erratic, suggesting that cycle-to-cycle variations in the period and/or the
noise of the data are large.  It is difficult to analytically determine the 
errors on 
Foster's wavelet Z-statistic, but a reasonable estimate of the {\it maximum}
error is the confusion limit of the spectrum, which can be measured as the 
half-width of the wavelet Z spectrum at each lag time.  For values of 
$c > 0.001$, the half-width of the peaks is much greater than 5 days, which is
more than the variation of the period maximum itself.  For values of 
$c = 0.00025$ and $0.001$, the half-widths are on the order of the observed 
variation.  Since this value is the confusion limit rather than the error in
period due to noise or other considerations, the actual error bars are likely
smaller.  Based on this, there appear to be marginally significant period 
changes in {\it o} Ceti at a level of 4-5 days over the course of the past 105
years.  Although the span of the data is short relative to the timescales in
question, there appears to be both a trend toward increasing average period,
and a marginally significant oscillation in period with a timescale
of $\sim 30-35$ years, and amplitude of 1-2 days.  This is a marginal
detection, but the increasing period is consistent with the expected 
evolutionary trend, and the oscillatory behavior is strikingly similar to 
that observed in the {\it meandering Miras} \citep{ZB02}, albeit with much 
lower amplitude.

\subsection{Multiperiodicity}

LSPs have been found in large numbers of long-period
semiregular variables in some surveys \citep{Derekas06,Soszynski07,Fraser08},
but as was mentioned in the introduction, the causes of these periods are 
generally unknown and they have not been conclusively detected in Miras.
\citet{BM97} noted that
several long periods were present in the {\it o} Ceti light curve
with marginal statistical significance, and one motivator for our
project was to investigate these claims with a longer and more complete
data set than that used in the earlier analysis.

When the entire 105 year light curve was analyzed together, no coherent
LSPs were detected in the data with any statistical significance.  This
suggests that such periods are not present, that they are not present 
throughout the entire light curve, or that they are at low amplitude.  To
test for transient LSPs, we subdivided the 105 year light curve into
6000 day segments, the length of which was chosen arbitrarily to cover 
more than a dozen cycles of the dominant variation, and which could in
principle detect periods as long as 3000 days.  The light curve was split
so that consecutive segments overlapped by 3000 days; any given point
in the light curve appears in two adjoining segments, but every other
segment represents a wholly independent data set.  All data segments were
again analyzed with the cleaning Fourier transform, and again no LSPs having
periods shorter than 3000 days were found with statistically significant
amplitude.  It is possible that there could be transient periods longer
than 3000 days in the full light curve, but there was no evidence for this
in the analysis of the full data set.

Finally, to search for any kind of quasiperiodic behavior or variations
with a characteristic timescale,
we employed an autoregressive moving average (ARMA) 
model to the data, followed by an autocorrelation analysis as devised by 
\citet{BJ76}.  This particular model includes a seasonal adjustment
(known as a SARIMA model), since it is already known that
the data contain a strong period of 333 days.
The raw visual observations were rebinned into 1 day averages, and
the data were resampled onto an even grid using the algorithm of
\citet{Reinsch67}.  We then formed a new set of data by subtracting both
the value of the nearest neighbor magnitude, as well as that of the
data point 333 days prior and its nearest neighbor:

\begin{equation}
W_{i} = (m_{i} - m_{i-1}) - (m_{i-333} - m_{i-334})
\end{equation}

\begin{figure}
\figurenum{4}
\label{figauto}
\epsscale{1.0}
\plotone{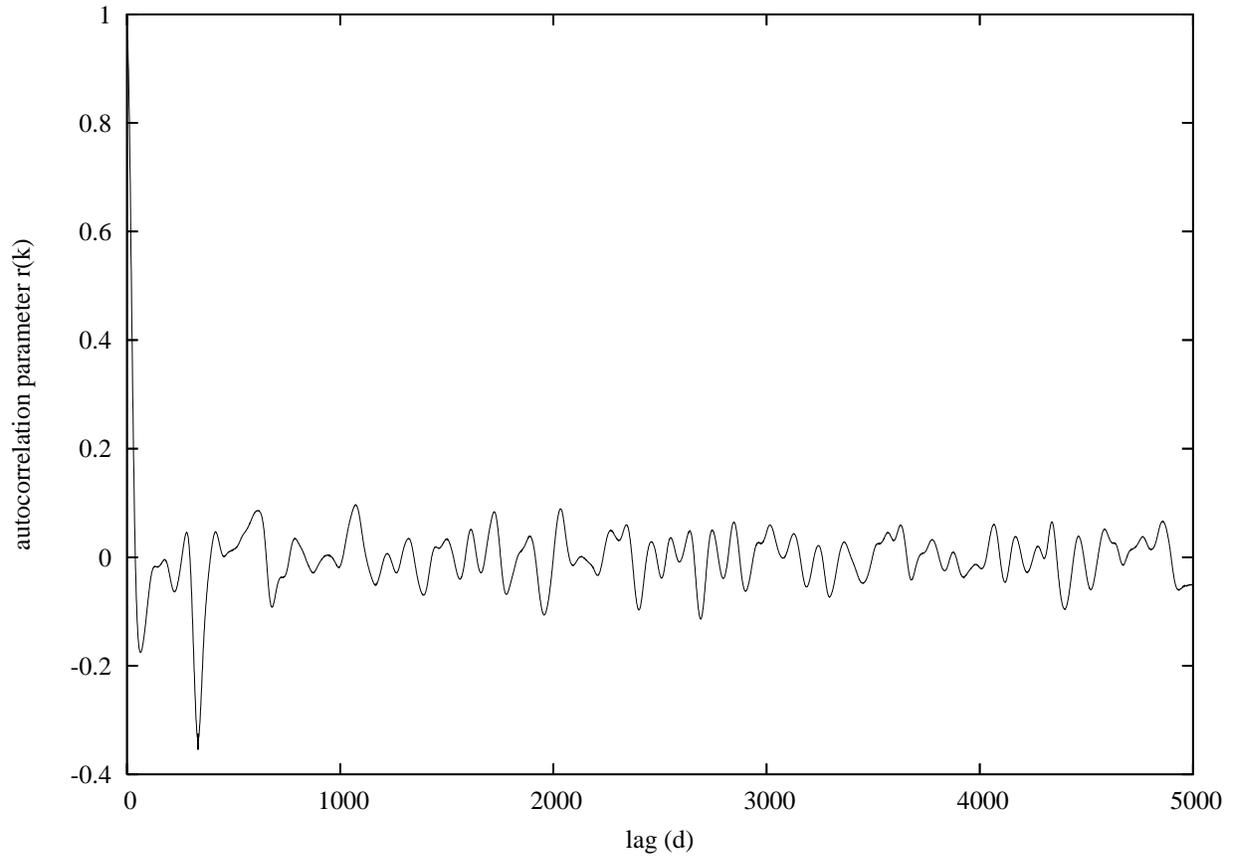}
\caption{Autocorrelation parameter $r(k)$ of the {\it o} Ceti data fit
with SARIMA model.  No significant correlations appear in the data following
the trend and seasonal adjustments, indicating there are no coherent LSPs
in the data.}
\end{figure}

This subtraction is completely analogous to the typical monthly or seasonal 
adjustment (e.g. $X_{i} - X_{i-12}$) often seen in similar autocorrelation 
analyses of terrestrial or socioeconomic data.  The new data set 
$(t_{i},W_{i})$ was
then analyzed with a \citet{BJ76} autocorrelation algorithm.  A plot of
the autocorrelation parameter $r(k)$ versus lag time for the adjusted data
is shown in Figure \ref{figauto}.  The adjusted data are essentially 
uncorrelated at any period, except for an anticorrelation at the known period
of the pulsation.  The lack of correlation peaks on any timescale up to
half the length of the light curve strongly suggests no periodic
or quasiperiodic variability is present in {\it o} Ceti.

As we will show in Section 3.4, the spectrum contains a strong red noise
component, with higher amplitude fluctuations at longer periods.  This
suggests that the peaks observed by \citet{BM97} may have been these 
stochastically generated fluctuations rather than the discrete modes
mentioned as a possible cause.  The amplitude of the background has a
$1/f^{\alpha}$ shape, with an amplitude of roughly 0.1 mag at
a period of 1000 days.  We can rule out the presence of LSPs in {\it o}
Ceti with amplitudes greater than this.  We cannot rule out very low
amplitude pulsations, but the stars exhibiting LSPs generally have amplitudes
larger than this limit.  

\subsection{Analysis of the 170 year light curve}

\begin{figure}
\figurenum{5}
\label{figoldlc}
\epsscale{1.0}
\plotone{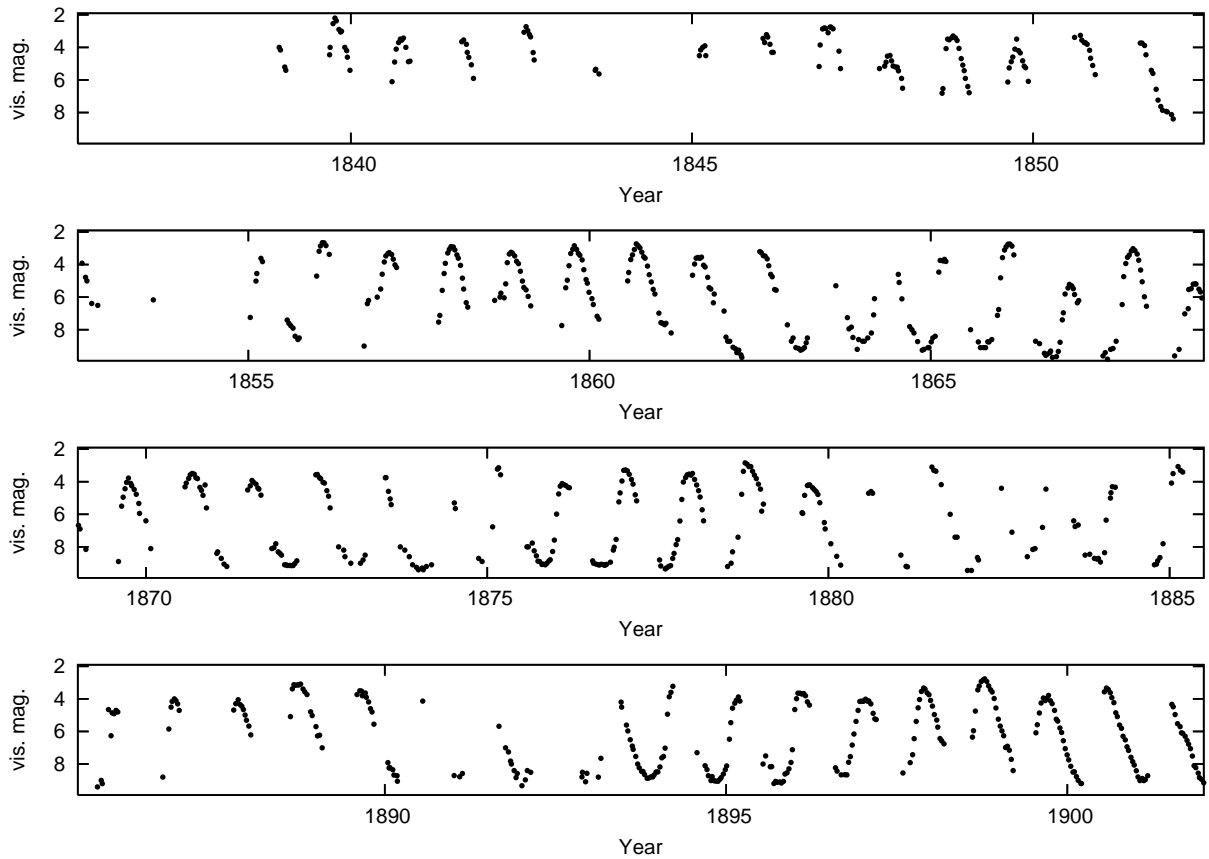}
\caption{Light curve of {\it o} Ceti, 1838-1902.  These data are much more
sparsely covered than the 105 year light curve shown in Figure 1, although
the overall behavior is clearly apparent.  Digitized data kindly provided 
by E. Zsoldos \& G. Marschalk\'{o}.}
\end{figure}

We performed some of the above analyses on a subset of the light curve kindly
provided by E.~Zsoldos and G.~Marschalko \citep{M04} to determine whether any
periodic behavior might be present during earlier times, or whether the use
of a longer set of data might raise the signal-to-noise ratio of 
low-amplitude variability in the 
spectrum.  The light curve provided by Marschalko contains data as far back
as the 17th century, but for the purposes of time-series analysis, only the
data beginning with the observations of F.~Argelander in the 1830s are 
well-sampled enough to provide reasonable coverage for our purposes, and
we used only data beginning at JD 2,392,722 (1838 December 13) up through
the start of our own data set at JD 2,416,000.  These data were combined,
and again averaged into 10 day wide bins.  The light curve of the early
data is shown in Figure \ref{figoldlc}.

\begin{figure}
\figurenum{6}
\label{figftcompare}
\epsscale{1.0}
\plotone{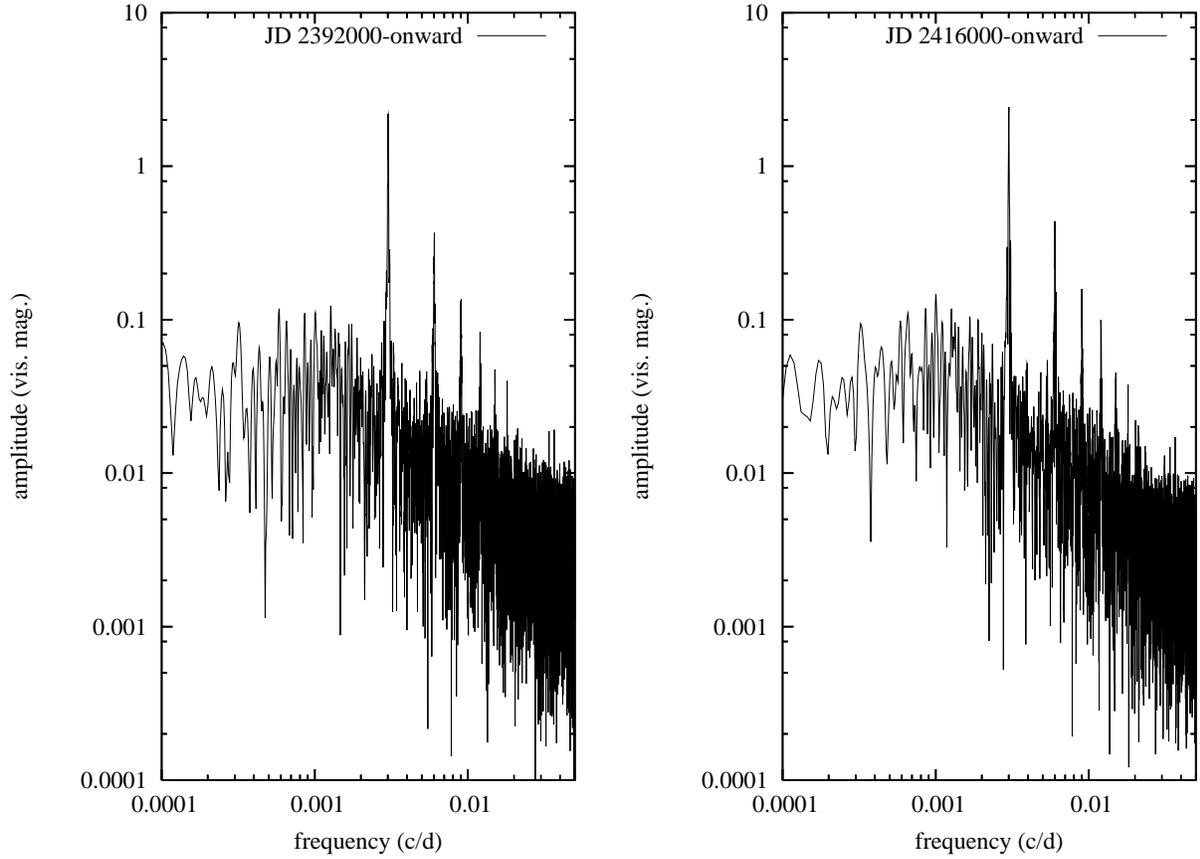}
\caption{Cleaned Fourier transforms of 105 year (right) and 170 year (left)
light curves.  Both light curves show the same overall features including
dominant frequencies and their amplitudes, as well as slope of the power
law, and flattening of the spectrum for frequencies below $0.001 c/d$.}
\end{figure}

\begin{figure}
\figurenum{7}
\label{figdpdt}
\epsscale{1.0}
\plotone{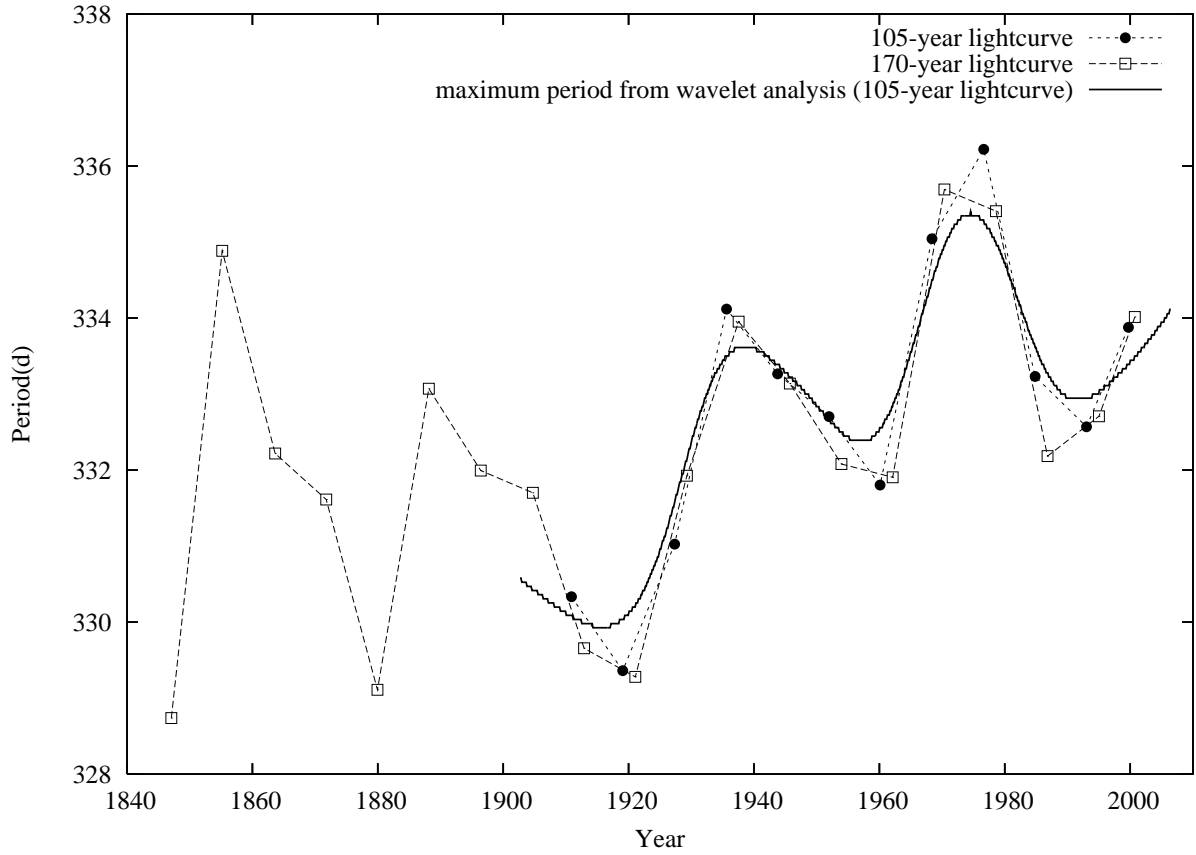}
\caption{Period change vs.~time as measured using 6000 day segments 
(points) and wavelet analysis (solid line).  The centroid periods of
the CLEANed spectra match those of the wavelet analysis, indicating
that the wavelet analysis is correctly measuring the correct period.
The oscillatory nature of the main period variation continues in the
longer, 170 year data set for data prior to 1900, suggesting that it 
is real.}
\end{figure}

When all of the data from JD 2,392,722 to the present are analyzed, the 
resulting spectrum (Figure \ref{figftcompare}) looks very similar to that of 
the light curve beginning on JD 2,416,000 as is expected.  There is no clear 
evidence of LSPs, and although the low-frequency noise structure is similar 
to that of the transform of the shorter light curve, the peaks do not always
match in frequency.  Thus we do not believe there are weak LSPs present in
the longer light curve.  We then subdivided the data into 6000 day segments
as was done for the shorter light curve, and again found the same variation
in period (Figure \ref{figdpdt}), with no evidence of transient LSPs.  We
note that the variations in period -- both the period increase and the
decades long modulation in period -- appear to extend back to the earliest
portion of the light curve.  However, we decided to discard the first six
thousand days of data (up to JD 2,398,000) because the amplitudes returned for
the dominant period were so discrepant that we believe the transform was
corrupted by the lack of observations near {\it o} Ceti's minimum.  The two
data points thus discarded do support the period modulation, but not the 
long-term trend.

\subsection{The low-frequency power law component}

\begin{figure}
\figurenum{8}
\label{figslope}
\epsscale{1.0}
\plotone{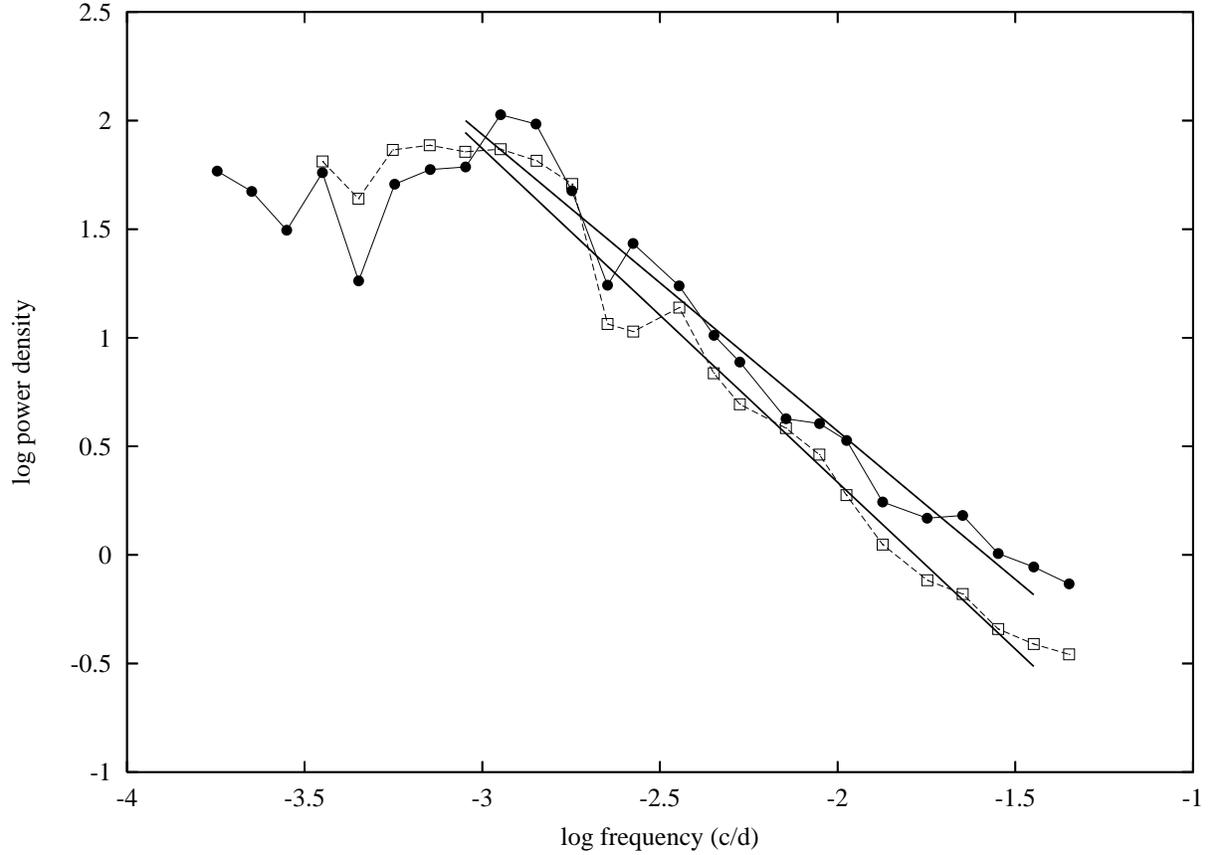}
\caption{Power density spectra of 105 year (open squares) and 170 year
(filled circles) light curves of {\it o} Ceti, showing the best-fitting 
power laws for $-3 < \log{f} < -1.4$.  The values of the power-law slopes
are $\alpha = -1.54$ for the 105 year light curve, and $\alpha=-1.37$
for the 170 year light curve, which are in reasonable agreement.  Both
spectra also show a turnover at $\log{f} < -3$, which suggests it is a
real effect, rather than due to the finite length of the shorter light
curve.}
\end{figure}

All of our time-series analyses indicated an excess of noise power at low
frequencies, indicative of an underlying red noise component to the variation
which is common in chaotic or quasiperiodic systems.  To investigate this 
feature, we followed the analysis of \citet{KSB06} to measure the spectral 
index of the spectrum.  We removed six bands from the spectrum directly rather
than prewhitening the data and Fourier transforming the residuals because the 
pulsation periods are not stable enough for effective prewhitening of such high
amplitudes.  The bands subtracted were defined by $i(f \pm \delta f)$, where
$i$=1,6, $f = 0.003$ c/d, and $\delta f = 0.0002$.  Following \citet{KSB06},
the spectrum was then averaged into logarithmic frequency bins 0.1 dex in size;
because the prefiltering removed large blocks of frequencies, the frequency of
any given bin was the average of the frequencies falling in that bin rather 
than the bin center.  The averaged power per bin was then normalized with
the length of the data set to calculate the power density.  We also performed
the same test on the longer, 170 year light curve.  The spectra of power 
density versus frequency for both light curves are shown in Figure 
\ref{figslope}, along with lines indicating the best-fit power laws for each.
For the 105 year light curve, the best-fit power law has $\alpha = -1.29$,
while for the 170 year light curve, the best-fit power law has 
$\alpha = -0.86$.  These two are different, but this has not taken into
account the fact that the slope of the 170 year data set includes more
low-frequency data lying within the turnover region.  When power laws are
fit to data only between $\log{f} = -3$ and $\log{f} = -1.4$, the slopes
for the 105 and 170 year light curves are $\alpha = -1.54$ and 
$\alpha = -1.37$ respectively, which agree within the uncertainties.

There is ambiguity in the nature of the power law which
leads to ambiguity in the physical explanation for the behavior.
A slope $\alpha = -1$ can approximate the slope over the entire range of 
the spectrum, but would indicate a power excess at $\log{f} \sim -3$.  A 
steeper slope of $\alpha = -1.4$ would fit the spectrum between 
$ -3 \leq \log{f} \leq -1.3$, but would indicate a strong turnover for 
$\log{f} < -3$.  Aside from the 
slight depression in power around $\log{f} \sim -2.5$ ($P \sim 333$ days),
the power density spectrum is a smooth continuum between $\log{f} \sim -3$
and $-1.5$, with a best-fit power law having a slope of $\sim -1.37$, and
a strong turnover for $\log{f} < -3$.  This could be interpreted as a cutoff
in power for low-frequency variability, either caused by the lack of 
variations on very long timescales, or a maximum amplitude for such 
variations.  If instead a single power law were used for the entire spectrum,
then there is evidence for a bump in power with periods around 1000 days.
We do not see coherent variations with those periods either in the full data
set or in the 6000 day segments, so we do not believe these could be unresolved
long-period modes.

\section{Discussion}

Our analysis of the visual light curve of {\it o} Ceti has shown three things: 
that the only definite {\it periodic} signal is that of the pulsation of 
{\it o} Ceti, that there is a marginally significant systematic variation in
period of approximately $\pm 2$ days over the past century, and that there
appears to be a low-frequency power-law component to the Fourier spectrum.
The use of a spectral cleaning algorithm that removes the effects of the
window function substantially reduced the number of peaks in the 
final spectrum, and what remained appears to be either directly related to 
the pulsation period and its integer harmonics, or to the underlying 
stochastic power law.  We set out to discover at the start of this project
whether {\it o} Ceti was among the AGB stars with LSPs,
and we have definitively ruled this out at the level of 0.05-0.1 mag
for periods between 300 and a few thousand days.

On the other hand, the computed power spectra of {\it o} Ceti -- both for 
the entire light curve and for the 3000 day data segments -- all show very
strong excess of power at low frequencies.  The physical cause of such a
power-law component is likely the brightness variations across the stellar
surface generated by convection.  The recent work of \citet{KSB06} clearly
showed that convection is strong in red supergiants, and the effect would
be similar if diminished for the smaller AGB stars.  We believe that the 
low-frequency spectrum in {\it o} Ceti is generated by the same processes that
cause the low-frequency power law in the red supergiants.  

For example, \citet{Sch75} predicted that convective
cells with sizes on the order of $1/4$ to $1/30$ of the visible portion of
the stellar disk may generate detectable photometric variations with
timescales on the order of the survival lifetime of the cell, namely a few
to several hundred days.  Likewise, \citet{ACN84} predict that large
convective cells in red giant envelopes appear to be those preferentially
selected.  The appearance, evolution, and destruction of these convective
cells (with the concurrent changes in local surface temperature, surface
brightness, and possibly dust opacity) would result in photometric variability.
If the cells are generated with a range of sizes as happens in stochastic 
processes like convection, it is reasonable that the resulting photometric
variability would have a power-law spectrum as well.  

In addition, we detect in {\it o} Ceti a weak feature showing an 
apparent flattening of the power law for periods longer than approximately 
1000 days.  The reality of this turnover and its cause are not
certain, and it may be due to the finite length of the light curve.  However,
the presence of the turnover in the much longer light curve including archival
data from the mid-19th century argues that the turnover is real.
One possible origin for a turnover is that the strong pulsation in {\it o} 
Ceti may cause dissipation of convective cells with turnover times longer than
the pulsation period of 333 days.  Another possibility is that the size of 
{\it o} Ceti itself, 
smaller than the red supergiants, may place an upper limit on the sizes
of convective cells that can form to those with turnover times less than
several hundred days.
Our results for {\it o} Ceti, and the results for the red supergiants obtained
by \citet{KSB06} 
highlight the necessity of high-angular resolution
observations of {\it o} Ceti and other Mira variables with the capability
of imaging the stellar surface variations.

\section{Conclusions}

Our time-series analysis of the longest continuous light curve of {\it o}
Ceti reveals no long-term, coherent periodic behavior at any period, to
a significance level of approximately 0.1 mag at $P = 1000$ days.
Our analysis has revealed the presence
of a low-frequency power-law component which can generate quasi-periodic, 
low-frequency variability on short timescales.  It is this low-frequency
power-law component that is responsible for the upper limit of 0.1 mag,
not the photometric precision of the visual data; if there is an LSP
present, it must have an amplitude lower than that of the low-frequency
stochastic variability.  We would have detected any LSP
with an amplitude greater than 0.05 mag if it were not overwhelmed by the 
intrinsic noise.  The physical origin of the stochastic variability 
is unknown, but it may be a signature of photometric
variability caused by convective supergranulation on large timescales first
suggested by \citet{Sch75}.  

We have detected a modulation in the dominant pulsation period of {\it o}
Ceti with a timescale of approximately 30 years.  This modulation is
detected with marginal statistical significance, but the timescale matches
other observed modulations in the physical behavior of the Mira AB system
\citep{Baize80,Karovska92}.

We have shown that using long-term visual light curves of Miras we can explore
low-frequency variability to amplitude limits below 0.1 mag.  A number
of Mira variables have very long light curves, and in principle periods of 
20 times the dominant pulsation period could be detected in many of them.
A large-scale population study of pulsating AGB stars looking specifically
for LSPs and correlating them with spectral type and
other stellar properties is warranted.  Our understanding of these objects
would be improved by careful scrutiny and analysis of the light curves 
themselves.

We wish to emphasize in closing that the significance limits on large ensembles
of visual data (for example the long-term Mira light curves) are clearly 
better than the 0.2-0.3 mag {\it per estimate} commonly quoted.  
When a large set of data is used in a study such as this, it is very 
straightforward to detect periodic behavior at amplitudes on the order of 
0.05 magnitudes or less.  Our nondetection of coherent, long-period
variability in {\it o} Ceti itself is not a limitation of the data quality,
but is caused by the presence of an intrinsic low-frequency power-law 
component which would overwhelm any existing coherent signal.  Visual 
observations 
{\it in sufficient quantity} may certainly provide much more sensitive 
measures of stellar variability than is assumed, and long spans of 
visual observations should be used to perform similar studies with other
bright and well observed Mira variables.

\acknowledgments

We thank the many thousands of observers around the world who have contributed 
observations of this and all other variable stars to the AAVSO and all other 
variable star organizations worldwide.  We extend our special thanks to 
R. Pickard of the BAAVSS and R. McIntosh of the RASNZ for personally 
supplying the data archives of those two organizations, and to the AFOEV 
and VSOLJ for making their archives available.  We thank E.~Zsoldos \& 
G.~Marschalk\'{o} for supplying an electronic copy of their
Mira light curve.  We also thank the anonymous referee whose comments
substantially improved this paper.  M.K. is a member of the 
{\em Chandra} X-ray Center, which is operated by the Smithsonian Astrophysical
Observatory under NASA Contract NAS8-03060.

\end{document}